# Detection and imaging of microcracking in complex media using coda wave interferometry (CWI) under linear resonance conditions

S. Toumi,[1,3] C. Mechri,[1,2] M. Bentahar [1], F. Boubenider [3] and R. El Guerjouma [1]

[1] *Université du Maine, Laboratoire d'Acoustique de l'Université du Maine, UMR CNRS 6613, Le Mans, France*

[2] *Centre de Transfert de Technologie du Mans, Le Mans, France*

[3] *Laboratoire de Physique des Matériaux, Ondes et Acoustique, USTHB, Alger, Algeria*

*This letter presents the development of ultrasonic coda wave interferometry to locate and image microcracks created within consolidated granular media, polymer concrete. Results show that microcracks can be detected through coda waves when the polymer concrete is submitted to a linear resonance. In addition, the multiple scattering revealed to be dependent on the plane in which the resonance is excited. In that sense, Acoustic emission measurements were performed under flexural resonances generated in (XY) and (XZ) planes to verify the existence of a possible structural anisotropy related to microcracks. Recorded acoustic signatures revealed important differences in the frequency contents, depending on the considered plane, with a consequent wealth on the involved mechanisms during the microcracks vibrations.*

The study of multiple scattering of elastic waves in complex media have started since more than half a century in geosciences, where the late part of the diffuse field was termed coda[1]. Being more sensitive to changes in the propagating medium than the single scattering waves, coda waves were used to monitor velocity variations of two successive signals, which are nearly alike, in the case of an earthquake [2] or an active source [3]. The term coda wave interferometry (CWI) was given by Snieder et al. where they proved the possibility of detecting weak velocity changes in solids using seismic and ultrasonic coda waves [4]. Later on, thermal CWI was applied on concrete and revealed to be of high sensitivity compared to time-of-flight techniques [5, 6]. At the time where techniques based on the intensity fluctuations created by isolated scatterers have proved limited detection capabilities [7], other techniques based on the linear interaction with defects revealed to be very sensitive to detect weak changes with a possibility to have their spatial extent in the medium[8-10]. On the other hand, stress-induced CWI experiments were performed to determine third order elastic constants of complex solids in the frame of the acoustoelastic theory[11]. Results revealed that in the case of a micro-cracked concrete the acoustoelastic coefficients increase as a function of the induced cracks[12]. The nonlinear interaction of the coda wave with defects was also determined in multiple scattering media through the nonlinear mixing with a pump wave[13]. However, one should keep in mind that the interaction of the coda with the micro-scatterers is not always sensitive enough and needs to be enhanced by increasing the dynamic strain level at the scatterers, for instance, through an increase of the dynamic excitation. Here one should pay attention to the fact that such a procedure, which is usually at the origin of the nonlinear behavior in many types of materials, is not always suitable especially in the case of consolidated granular media. Indeed, conditioning and relaxation effects might appear even in the absence of micro-cracks [14] and disturb the defect characterization procedure. In this work, we present a method to predispose micro-cracks to interact with a high frequency wave when the propagating medium is submitted to a linear vibration. The method allows detecting, locating and imaging defects using an original approach very different form the passive imaging technique, which is generally hard to implement when real time monitoring is necessary.

The characterized consolidated granular material, whose dimensions are 160x40x11 mm$^3$, consists of an epoxy resin matrix reinforced by sand and aggregates at 40%, 30%, and 30% volume fraction, respectively. This polymer concrete was submitted to a three-point bending fatigue test performed under 3 kN loading force, where the distance between the supporting pins was set at 120 mm. The created microchanges in such a complex medium can be detected by taking advantage of the multiply scattered waves. Structural variations are determined by cross-correlating two waveforms taken before and after the fatigue test or at low and high excitation amplitudes. The time-windowed normalized crosscorrelation function $R(t)$ is expressed as:

$$R(t_s) = \frac{\int_{t-T}^{t+T} u(t')\tilde{u}(t'+t_s)dt'}{\sqrt{\int_{t-T}^{t+T} u^2(t')dt' \int_{t-T}^{t+T} \tilde{u}^2(t'+t_s)dt'}} \qquad (1)$$

$u(t)$ is the waveform corresponding to the initial state (or low excitation) and $\tilde{u}(t)$ is the waveform obtained at fatigued state (or high excitation). $2T$ is the length of the considered time window, which is centered around $t$ ( $t \gg T$ ). Consequently, the decorrelation $K(t_s)$ between $u(t)$ and $\tilde{u}(t)$ can be determined as: $K(t_s) = 1 - R(t_s)$

To detect and locate the created scatterers (microcracks) within the weakly fatigued polymer concrete, the application of the CWI for different ultrasonic paths did not show any clear time delay between $u(t)$ and $\tilde{u}(t)$ even for increasing excitation amplitudes, where the correlation function produced a sequence of autocorrelations i.e. $K(t_s) \cong 0$. In order to increase the sensitivity of the CWI, we used a shaker controlled by a power amplifier (46 dB) delivering a peak-to-peak excitation up to 120 V (see Fig.1). Vibration modes of the polymer concrete sample generated in a clamped-free configuration are detected using an accelerometer attached to the free side of the sample. Since the aim of these experiments is not to create nonlinear resonances, we performed preliminary fast and slow nonlinear dynamics measurements in order to probe the conditioning of the consolidated granular sample and its subsequent relaxation. For the three first flexural resonances, results show that the material is vibrating in the linear regime as long as the excitation amplitude is kept below 20 mV before amplification. Under the linear vibration conditions, we perform through transmission measurements of ultrasonic pulses generated using identical large band transducers mounted opposite each other, where the emitter is excited with a burst signal at 440 kHz. In order to generate the resonance in the linear regime, the shaker was excited continuously at 10 mVpp while the ultrasonic emitter transducer was excited by a pulse generator adjusted to deliver a 50 mVpp amplified at 46 dB. The experimental setup was calibrated through a set of measurements, where different aspects like positioning of sensors and coupling were taken into account. The numerous measurements showed that for a given position of the transducer, the reproducibility of the delay between the multiply scattered signals does not exceed 50 ns (which correspond to the sampling period of the acquisition system). On the other hand, due to the sensitivity of the coda wave to the environmental conditions (that may change between two measurements), reference signals were recorded at every position when the material is at rest.

Evolution of the decorrelation coefficient as a function of the transducers positions along the x-axis is presented in Fig.2. The latter shows that depending on the excited linear resonance, the coda of the ultrasonic signals is affected differently depending on the generated strain at the scatterers. Furthermore, Fig.2 shows that microcracks created during the fatigue test do not seem to propagate along a single line but are diffused between the supporting pins. As a function of the excited resonance, the strain distribution at the microcracks changes with a consequent impact on the coda of the recorded signals observed through the decorrelation coefficient $K$. This result is in accordance with previous works on the fracture behavior of inhomogeneous materials submitted to a three point bending test. In such materials, microcracks result from the interaction between the gradient of the stress field and the distribution of the fracture stresses, where the stress can be higher than the lower limit of the fracture distribution even at the end of the linear domain [15,16].

In the view of imaging the scattering of the ultrasonic waves, contact transducers were replaced with air-coupled ultrasonic transducers whose frequency bandwidth goes from ~ 300 kHz to ~ 700 kHz. In addition of being contactless, the air-coupled method allows improving the spatial resolution and is much faster than the contact approach. Air-coupled transducers are excited with a sine-Gaussian profile signal emitted with 1 kHz repetition frequency. Received signals are sampled at 15 MHz over a dynamic range of 16 bits and amplified at 60 dB. Due to the important difference existing between the acoustic impedances of the air ($Z_{air} \cong 416 Rayl$) and the polymer concrete ($Z_{PC} \cong 9 MRayl$), the reflection coefficient of the ultrasonic wave is important $R \cong 99.98\%$. However, the high voltage used in the excitation amplitude (~ 200 V) associated to an averaging of the received signals (~ 50) allowed having a good signal-to-noise ratio (SNR), which is around 30 dB for the ballistic wave. The SNR corresponding to the coda of the recorded signals was above 12 dB. Under these conditions, reference signals correspond to the propagation through the polymer concrete before activating the shaker. Received signals are then recorded under weak vibrations, as explained above, and coda waves are analyzed by considering a time window corresponding to ~20 μs (8 periods). Such a procedure allows getting a through transmission imaging of the polymer concrete as shown in Fig. 3. The latter shows that under a linear resonance, the sensitivity of the CWI to micro-cracks is improved and allows an active and simple imaging of the microcracked area using an original approach.

When the same air-coupled coda experiments are repeated for resonances excited in the XY plane, decorrelation coefficients revealed to be small ($K(t_s) < 0.02$). Such a weak interaction between the ultrasonic wave and the existing microcracks is mainly due to the direction along which the mechanical force is applied during the bending test. The moderate force load creates micro-cracks within the polymer matrix and/or the interface between aggregates and matrix. The loading conditions make the cracks mainly in the force direction [17]. However, one might expect important cracks kinking angles away from the beam center with a local influence of the aggregates size and distribution [16]. In view of the cracks structural anisotropy, it is expected that mechanisms at the microcracks are changing depending on the considered resonance plane. In order to verify the change in the generated mechanisms, acoustic emission (AE) measurements were performed when micro-cracked samples are submitted to the same flexural resonances in the XY and XZ planes. Signals were collected with 5 MHz sampling rate along 5012 points for each AE hit using broadband piezoelectric sensors, where signals are 40 dB pre-amplified. First, we noticed that the acoustic activity in terms of the generated number of signals is qualitatively not the same. Indeed, the acoustic activity was found to be more important in the XZ plane as can be seen on Fig. 4. In addition, the characteristics of the acoustic emission signals are not the same. This has been verified through the frequency contents of the acoustic emission signals recorded along one resonance cycle in both configurations. Note that the frequency range of the acoustic emission signals is completely different from the ones used to generate the different resonances, which are below 4 kHz. In the same figure we can see that microcracks behave differently depending on the plane in which resonances are excited. Indeed, in view of their orientation, microcracks can be submitted to shear and/or compression forces, which can enhance mechanisms as different as clapping and sliding involving different frequency components. At the time when we think that changing a resonance plane will mainly favor one mechanism over another, we believe that the understanding of the experimental observations need a deeper analysis. Indeed, the presence of memory effects beside the aforementioned activated mechanisms (clapping or sliding) makes the analytical formulation of the relationship between stress and strain not feasible. As an alternative, a multi-state statistical description of the microcracked polymer concrete, based on the generalized Preisach–Mayergoyz (PM) formalism, seems to be a promising approach in order to link the behavior of simple mesoscopic elements (at the microscopic scale to describe the microcracks behavior) to the observations performed at the macroscopic scale [17]. Our future work will be developed in this direction.

**Figure captions**

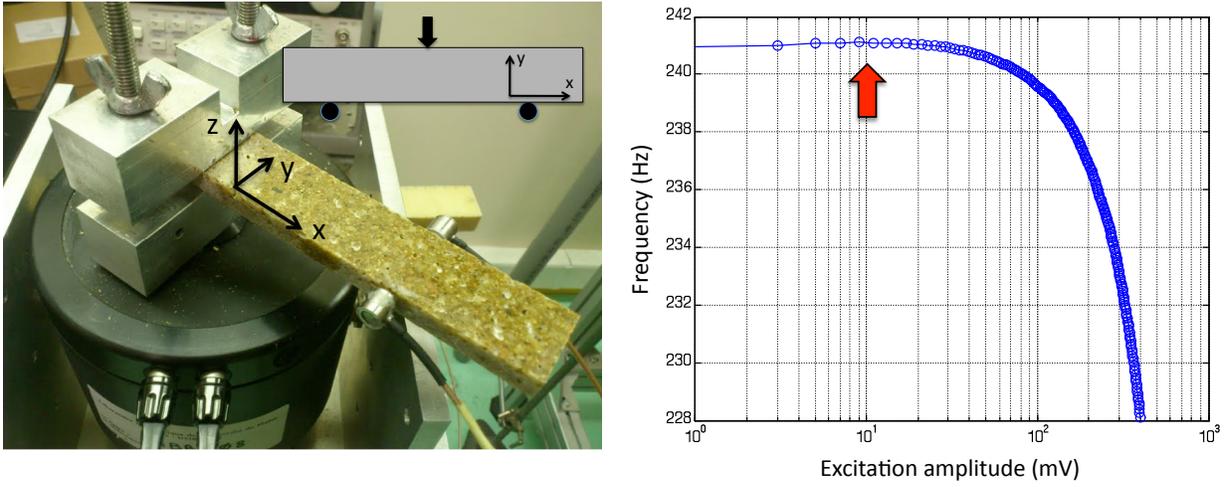

FIG. 1. (Color online) Left: The polymer based concrete sample mounted on the shaker where coda waves are detected at different positions along x-axis using two identical ultrasonic transducers: note that the three-point bending is performed in the XY plane. Right: Resonance frequency change corresponding to the fundamental bending resonance under an increasing excitation: the arrow shows the excitation at which linear resonances are excited (Right).

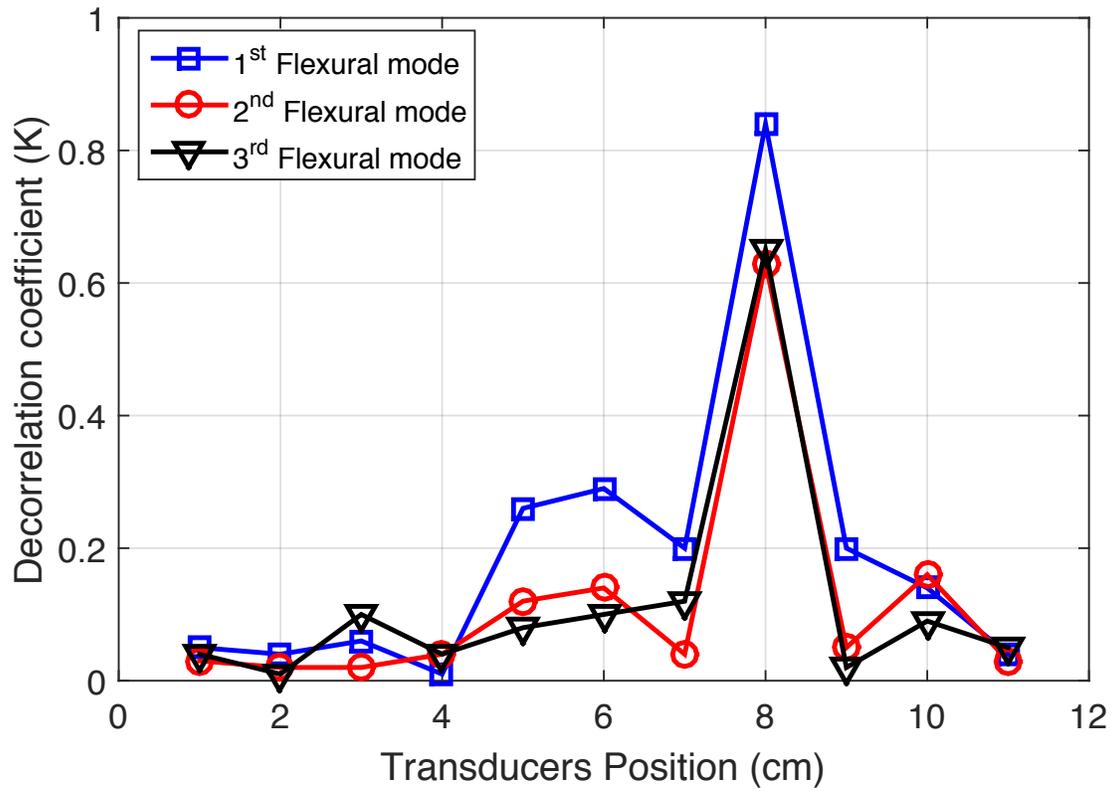

FIG. 2. (Color online) Decorrelation coefficient K determined at different positions along X-axis when the polymer concrete is submitted to a linear flexural resonance. Results show that microcracks distribution is not symmetric to the middle of the concrete beam (around 8cm)

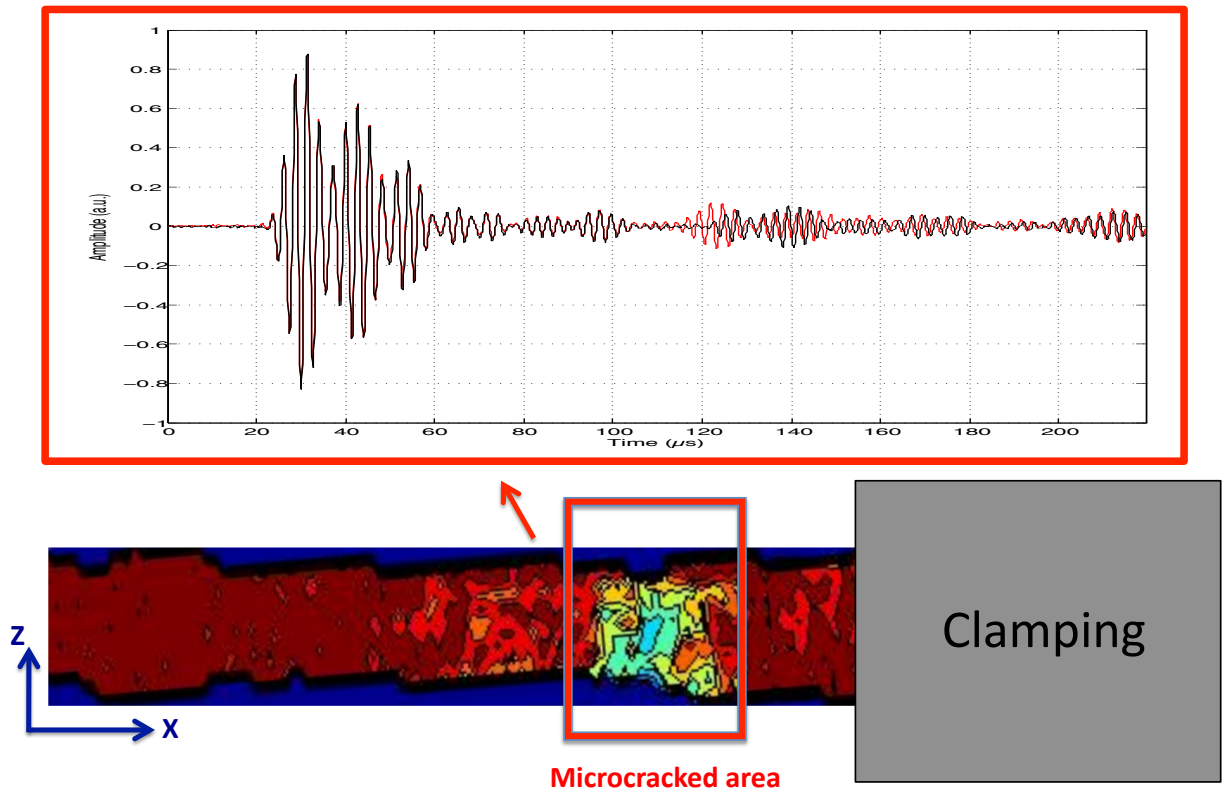

FIG. 3. (Color online) Air-coupled ultrasonic imaging of the microcracks created within the polymer concrete. Results are obtained when the concrete beam is excited at the 1[st] flexural resonance in the XZ plane. Equivalent images were found for the 2[nd] and 3[rd] flexural resonances. The A-Scan signal is an example taken from the microcracked area.

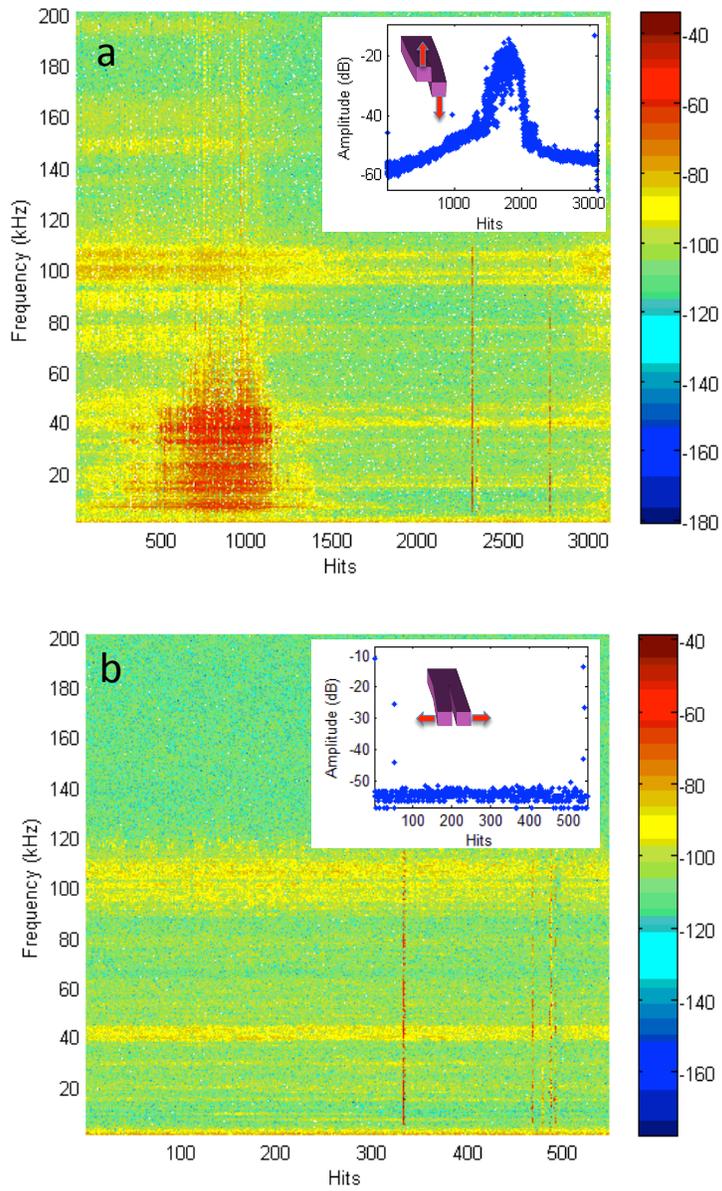

FIG. 4. (Color online) Frequency contents of the recorded acoustic emission signals along one resonance cycle in the XZ (a) and XY (b) planes, respectively. Figures show the possibility to study the mechanisms (the main ones) generated at the microcracks using the characteristics of the recorded acoustic emission hits.